\documentclass[twocolumn,showpacs,superscriptaddress,amsmath,amssymb]{revtex4}

\usepackage{float,epsfig}

\begin{document}

\title{Low-momentum nucleon-nucleon interaction and shell-model calculations}

\author{A. Covello} 
\author{L. Coraggio} 
\author{A. Gargano} 
\author{N. Itaco}
\affiliation{Dipartimento di Scienze Fisiche, Universit\`a
di Napoli Federico II, \\ and Istituto Nazionale di Fisica Nucleare, \\
Complesso Universitario di Monte  S. Angelo, Via Cintia - I-80126 Napoli,
Italy}

\date{\today}

\begin{abstract}
We discuss the use of the low-momentum nucleon-nucleon ($NN$) interaction 
$V_{\rm low-k}$ in the derivation of the shell-model effective interaction and emphasize its practical value as an alternative to the Brueckner $G$-matrix method. We present some selected results of our current study of exotic nuclei around closed shells, which have been obtained starting from the CD-Bonn potential. We also show some results of calculations performed with different phase-shift equivalent $NN$ potentials, and discuss the effect of changes in the cutoff momentum which defines the $V_{\rm low-k}$ potential.
\end{abstract}

\maketitle

\section{Introduction}

In recent years, shell-model calculations employing realistic effective interactions derived from modern nucleon-nucleon ($NN$) potentials  have produced results in remarkably good agreement with experimental data for a number of nuclei in various mass regions. A main difficulty encountered in this kind of calculations is the short-range repulsion contained in the bare $NN$ potential $V_{NN}$, which prevents its direct use in nuclear structure calculations. This difficulty is usually overcome by resorting to the well-known Brueckner $G$-matrix method.

Recently, a new approach has been proposed \cite{bogner02,kuo02} which consists in deriving from $V_{NN}$ a low-momentum potential $V_{\rm low-k}$ defined within a given cutoff momentum $\Lambda$. This is a smooth potential which can be used directly to derive the shell-model effective interaction $V_{\rm eff}$. We have performed shell-model calculations \cite{bogner02,covello02,covello03} using this method as well as the traditional $G$-matrix one. Comparison of the results has shown that the former provides an advantageous alternative to the latter.

The main purpose of this paper is to give a brief survey of the practical application of
the $V_{\rm low-k}$ approach in realistic shell-model calculations. In section 2 we give an outline of the derivation of $V_{\rm low-k}$ and $V_{\rm eff}$,
while in section 3 we show a few selected results of our current study of nuclei around doubly magic $^{132}$Sn, which have been obtained starting from the CD-Bonn 
$NN$ potential \cite{machleidt01}. In section 4 we present some preliminary results of a study performed with different phase-shift equivalent $NN$ potentials and examine their dependence on the choice of $\Lambda$.   

%\vfill\eject

\section{Outline of theoretical framework}

As pointed out in the Introduction, we ``smooth out" the strong repulsive core contained in the bare $NN$ potential $V_{NN}$ by constructing a low-momentum  potential
$V_{\rm low-k}$. This is achieved by integrating out the high-momentum modes of $V_{NN}$ down to a cutoff momentum  $\Lambda$. This integration is carried out with the requirement that the deuteron binding energy and low-energy phase shifts of $V_{NN}$ are preserved by $V_{\rm low-k}$. This requirementy may be satisfied by the following $T$-matrix equivalence approach. We start from the half-on-shell $T$ matrix for $V_{NN}$ 
\begin{eqnarray}
T(k',k,k^2) = V_{NN}(k',k) + \wp \int _0 ^{\infty} q^2 dq
V_{NN}(k',q) \nonumber \\ 
\frac{1}{k^2-q^2} T(q,k,k^2 ) ~~,~~~~~~~~~~~~~~~~~~~
\end{eqnarray}

where $\wp$ denotes the principal value and  $k,~k'$, and $q$ stand for the relative momenta. 
The effective low-momentum $T$ matrix is then defined by
\begin{eqnarray}
T_{\rm low-k }(p',p,p^2) = V_{\rm low-k }(p',p) \wp \int _0 ^{\Lambda}
q^2 dq  V_{\rm low-k }(p',q) \nonumber \\
\frac{1}{p^2-q^2} T_{\rm low-k} (q,p,p^2) ~~,~~~~~~~~~~~~
\end{eqnarray}

where the intermediate state momentum $q$ is integrated from 0 to the momentum space cutoff $\Lambda$ and $(p',p) \leq \Lambda$. 
The above $T$ matrices are required to satisfy the condition 
\begin{equation}
T(p',p,p^2)= T_{\rm low-k }(p',p,p^2) \, ; ~~ (p',p) \leq \Lambda \,.
\end{equation}
\noindent 
The above equations define the effective low-momentum interaction $V_{\rm low-k}$, and it has been shown \cite{bogner02} that they are satisfied by the solution:
\begin{equation}
V_{\rm low-k} = \hat{Q} - \hat{Q'} \int \hat{Q} + \hat{Q'} \int \hat{Q} \int
\hat{Q} - \hat{Q'} \int \hat{Q} \int \hat{Q} \int \hat{Q} + ~...~~,
\end{equation}
which is the well known Kuo-Lee-Ratcliff (KLR) folded-diagram expansion \cite{KLR71,kuo90}, originally designed for constructing  shell-model effective interactions.
In the above equation $\hat{Q}$ is an irreducible vertex function whose intermediate states are all beyond $\Lambda$ and $\hat{Q'}$ is obtained by removing from $\hat{Q}$ its terms first order in the interaction $V_{NN}$. In addition to the preservation of the half-on-shell $T$ matrix, which implies preservation of the phase shifts, this $V_{\rm low-k}$ preserves the deuteron binding energy, since eigenvalues are preserved by the KLR effective interaction. 
For any value of $\Lambda$, the low-momentum effective interaction of equation (4) can be calculated very accurately using iteration methods. Our calculation of $V_{\rm low-k}$ is performed by employing the iteration method proposed in \cite{andreozzi96}, which is based on the Lee-Suzuki similarity transformation \cite{suzuki80}. 

The $V_{\rm low-k}$ given by the $T$-matrix equivalence approach mentioned above is not Hermitian. Therefore, an additional transformation is needed to make it Hermitian. To this end, we resort to the Hermitization procedure suggested in
\cite{andreozzi96}, which makes use of the Cholesky decomposition of symmetric positive definite matrices. In a very recent paper \cite{holt04} it has been shown that the $V_{\rm low-k}$ obtained through this method belongs to a family of Hermitian low-momentum $NN$ interactions which preserve the deuteron binding energy and are phase shift equivalent, all reproducing empirical phase shifts up to the cutoff momentum 
$\Lambda$. From a comparison of shell-model matrix elements calculated with Hermitian 
$V_{\rm low-k}$'s given by various methods it has turned out \cite{holt04} that they are approximately equivalent to each other. 

Having outlined the derivation of the low-momentum interaction $V_{\rm low-k}$ from 
$V_{NN}$, we turn now to shell-model calculations in which we use $V_{\rm low-k}$ as the input interaction. We derive the model space effective interaction 
$V_{\rm eff}$ by employing a folded-diagram method, which has been successfully applied  to many nuclei \cite {covello01} using $G$-matrix interactions. Since $V_{\rm low-k}$ is already a smooth potential, it is no longer necessary to calculate the $G$ matrix. We therefore perform shell-model calculations following the same procedure as described, for instance, in \cite{jiang92,covello97}, except that the
$G$ matrix used there is replaced by $V_{\rm low-k}$. More precisely, we first calculate the 
$\hat{Q}$-box including diagrams up to second order in $V_{\rm low-k}$. The shell-model effective interaction is then obtained by summing up the $\hat{Q}$-box folded diagram series using the Lee-Suzuki iteration method \cite{suzuki80}.

\section{Results and comparison with experiment}

Shell-model studies of neutron-rich nuclei in the close vicinity to doubly magic $^{132}$Sn  offer the opportunity to test the matrix elements of the effective interaction well away from the valley of stability.  We have recently 
studied \cite{covello05} several of these nuclei using as input interaction the $V_{\rm low-k}$ derived from the CD-Bonn $NN$ potential. We report here some selected results of our current study of nuclei with valence neutrons in the 82-126 shell, focusing attention on $^{134}$Sn. This is the heaviest Sn isotope for which some experimental information on excited states is presently available. 

In our calculation we let the two valence neutrons occupy the six single-particle levels $0h_{9/2}$, $1f_{7/2}$, $1f_{5/2}$, $2p_{3/2}$, $2p_{1/2}$, and $0i_{13/2}$ of the 82-126 shell. 
As regards the single-neutron energies, they have been taken  from the experimental spectrum of $^{133}$Sn, except that relative to the $i_{13/2}$ level which has not been observed. The latter has been taken from Ref. \cite{coraggio02b}, where all the adopted values are also reported. In this connection, it should be mentioned that the calculations performed in this previous work differ from the present ones mainly because we used there the $G$-matrix formalism.
The results presented in this section have all been obtained with a $V_{\rm low-k}$ derived from the CD-Bonn $NN$ potential and confined within a cutoff momentum $\Lambda$ = 2.1 fm$^{-1}$. This value has been chosen according to the criterion discussed in  \cite{bogner02}.

\begin{figure}[H]
\begin{center}
\includegraphics[scale=0.8,angle=0]{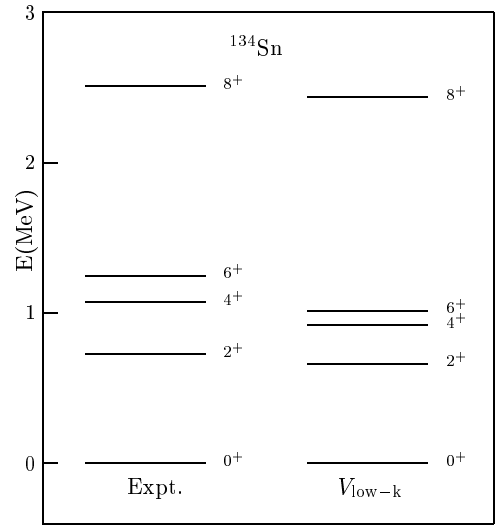}
\end{center}
\caption{Experimental and calculated spectrum of $^{134}$Sn.}
\end{figure}

From figure 1 we see that the calculated spectrum of $^{134}$Sn is in good agreement with the existing experimental data \cite{zhang97,korgul00}.
In figure 2 we compare with experiment our values of the excitation energy of the $2^+$ state  for the three isotopes $^{128}$Sn, $^{130}$Sn and $^{134}$Sn. The experimental value for $^{132}$Sn is also reported to show the trend across the $N=82$ magic number. The neutron single-hole energies used in the calculations for  $^{128}$Sn and $^{130}$Sn are the same as those adopted in our recent study \cite{genevey03} of $^{129}$In and $^{129}$Sb. 

Very recently \cite{beene04}, the $B(E2;0^+ \rightarrow 2_1^+)$ values in  $^{132}$Sn and $^{134}$Sn have been measured using Coulomb excitation of neutron-rich radioactive ion beams. We have calculated this $B(E2)$ for $^{134}$Sn with an effective neutron charge of $0.70e$, according to our early 
study \cite{coraggio02b}. The experimental and theoretical values are shown in figure 3, where we also compare with 
experiment \cite{radford03} our $B(E2;0^+ \rightarrow 2_1^+)$ values for $^{128}$Sn and $^{130}$Sn calculated with an effective neutron-hole charge of $0.78e$ \cite{genevey03}. 

From figures 2 and 3 we see that our calculations reproduce remarkably well the trend of  both the first excited  state and the $B(E2;0^+ \rightarrow 2_1^+)$ across the $N=82$ magic number.

\begin{figure}[H]
\begin{center}
\includegraphics[scale=0.45,angle=0]{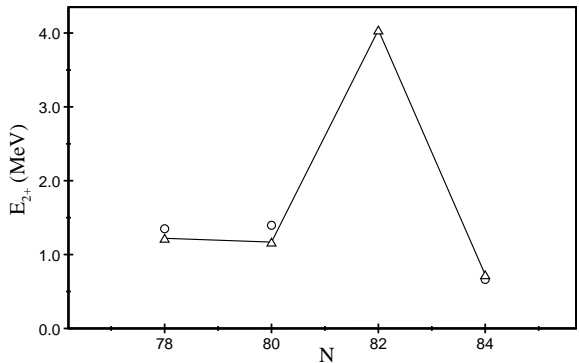}
\end{center}
\caption{Energy of the first $2^+$ excited state in Sn isotopes around $^{132}$Sn. The calculated results are represented by circles and the experimental data by triangles.}
\end{figure}

\begin{figure}[H]
\begin{center}
\includegraphics[scale=0.45,angle=0]{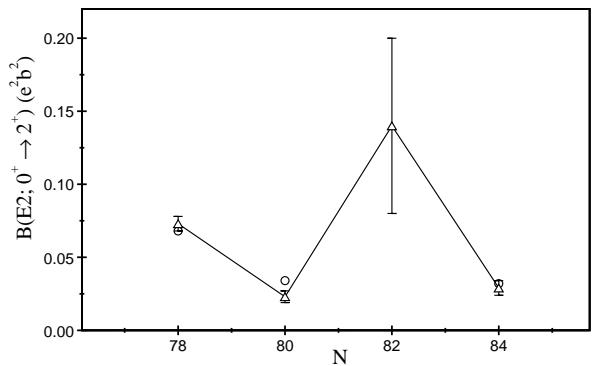}
\end{center}
\caption{Experimental and calculated values of $B(E2;0^+ \rightarrow 2^+_1)$
for Sn isotopes around $^{132}$Sn.}
\end{figure}

\section{Phase-shift equivalent nucleon-nucleon potentials and shell-model calculations}
The $NN$ potential $V_{\rm low-k}$ is currently attracting much attention. On the one hand, it is being applied in various contexts, for instance in the study of few-body
problems \cite{fujii04,nogga04}. On the other hand, its properties are being studied in detail, in particular as regards the dependence on the original potential model. 
In fact, there is evidence \cite{bogner03,bogner03b,coraggio05} that $V_{\rm low-k}$'s derived from different  phase-shift equivalent potentials, such as CD-Bonn, 
Nijmegen \cite{stoks94}, Argonne $v_{18}$ (AV18) \cite{wiringa95}, are nearly identical.

We are currently investigating \cite{covello05} the dependence of nuclear structure results on the $NN$ potential used to derive the shell-model effective interaction through the $V_{\rm low-k}$ approach.
Here we present some preliminary results obtained for the nucleus with two valence 
protons $^{134}$Te, which provides a good testing ground for this investigation.
We consider the three phase-shift equivalent 
$NN$ potentials NijmegenII, AV18 and CD-Bonn. As regards the cutoff $\Lambda$,
we let it vary from 1.5 to 2.5 $\rm fm^{-1}$.  In all cases we compare the calculated spectrum of $^{134}$Te with the experimental one up to about 4.5 MeV excitation energy
(this includes sixsteen excited states) and calculate the $rms$ deviation
$\sigma$\footnote{We define $\sigma =\{(1/N_{d}) \, \Sigma_{i} \, [E_{exp}(i)-E_{calc}(i)]^{2}\}^{1/2}$, where $N_{d}$ is the number of data points.}. 

\begin{figure}[H]
\begin{center}
\includegraphics[scale=0.5,angle=0]{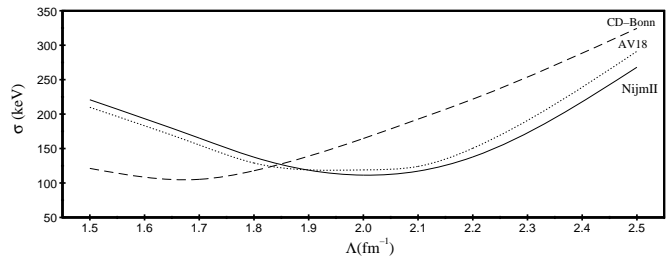}
\end{center}
\caption{Behavior of the {\it rms} deviation $\sigma$ relative to the spectrum of $^{134}$Te as a function of $\Lambda$ for different $NN$ potentials. See text for details.}
\end{figure}

In figure 4 we show the behavior of $\sigma$ as a function of $\Lambda$ for the three potentials. We see that the curves relative to NijmegenII (NijmII) and AV18
practically overlap each other while that for the CD-Bonn potential has a rather different pattern. In particular, the minimum of $\sigma$ for CD-Bonn is at 
$\Lambda \sim 1.7 \, \rm fm^{-1}$ while for the other two potentials it lies at
$\Lambda \sim 2 \, \rm fm^{-1}$. The minimum value of $\sigma$ for the three potentials is however almost equal and also the energies of the various states are practically the same. By way of illustration, we report in table 1  the ground-state energy of $^{134}$Te (relative to doubly magic $^{132}$Sn) and the excitation energies of the first three 
positive-parity yrast states.

\begin{table}
\caption{Energy levels (in MeV) of $^{134}$Te . Predictions by different $NN$ potentials are  compared with  experiment.}
\begin{ruledtabular}
\begin{tabular}{lcccc}
  & NijmII & AV18  & CD-Bonn & Expt\\
 $E_{gs}$       & -20.793 & -20.778 & -20.778 & -20.560\\
 $E(2^{+})$      & 1.342 & 1.325 & 1.315 & 1.279\\
 $E(4^{+})$      & 1.611 & 1.599 & 1.603 &1.576\\
 $E(6^{+})$      & 1.741 & 1.726 & 1.742 &1.691\\
\end{tabular}
\end{ruledtabular}
\end{table}

From figure 4 it also appears that all three curves are rather flat around the minimum. As a consequence, the quality of agreement between theory and experiment
does not change significantly for moderate changes in the value of $\Lambda$.
More precisely, for the CD-Bonn potential $\sigma$ remains below 150 keV for values of 
$\Lambda$ between 1.5 and 1.9 fm$^{-1}$ while for the other two potentials this occurs for $\Lambda$ between 1.8 and 2.2 fm$^{-1}$. 

The main conclusion of this preliminary study is that, allowing for limited  changes in the value of $\Lambda$, nuclear structure results obtained from 
$V_{\rm low-k}$'s extracted from different $NN$ potentials are practically independent of
the input potential. It therefore appears that the low-momentum interaction $V_{\rm low-k}$ gives an approximately unique representation of the $NN$ potential. This is quite in line with the conclusions of \cite{holt04b}. 

At this point, a comment on the results presented in this section is in order. As mentioned in section 2, we include in the $\hat{Q}$-box all diagrams up to second order in 
$V_{\rm low-k}$. In the calculation of these diagrams we have inserted intermediate states composed of particle and hole states restricted to the two major shells above and below the Fermi surface. However, as a development of our study, we are currently investigating the effect of increasing the number of intermediate states. The first results indicate that the minimum value of $\sigma$ occurs at values of $\Lambda$ which are somewhat larger than those reported here. Our final findings will be the subject of a forthcoming publication.  
 
\begin{acknowledgments}
This work was supported in part by the Italian Ministero dell'Istruzione, dell'Universit\`a e della Ricerca (MIUR).
\end{acknowledgments}

\end{document}